\documentclass{article}

\usepackage{amsmath}
\usepackage{amssymb}
\usepackage{theorem}
\usepackage{stmaryrd}
\usepackage{pstricks}
\usepackage{pst-node}
\usepackage{epsfig}
\usepackage{algorithm}
\usepackage{algorithmic}
\usepackage{bbold}
\usepackage{subfigure}

\usepackage{snapshot}

\def\Qed{\hfil\vrule width1em height0pt depth0pt\nobreak
\hfill $\blacksquare$\vskip\baselineskip}

\newtheorem{theorem}{Theorem}[section]
\newtheorem{proposition}[theorem]{Proposition}
\newtheorem{lemma}[theorem]{Lemma}

\newtheorem{definition}[theorem]{Definition}

\newenvironment{proof}%
{{\slshape \textbf{Proof: }}}
{\Qed}

\newcommand{\Thmref}[1]{Theorem~\ref{thm:#1}}
\newcommand{\Propref}[1]{Proposition~\ref{prop:#1}}
\newcommand{\Eqref}[1]{Equation~\ref{eq:#1}}

\begin{document}

\title{Duality of Fix-Points for Distributive Lattices}

\author{Prahladavaradan Sampath}
\maketitle

\begin{abstract}
  We present a novel algorithm for calculating fix-points. The
  algorithm calculates fix-points of an endo-function $f$ on a
  distributive lattice, by performing reachability computation a graph
  derived from the dual of $f$; this is in comparison to traditional
  algorithms that are based on iterated application of $f$ until a
  fix-point is reached.
\end{abstract}

\section{Introduction}\label{sec:introduction}
In this paper we cast the problem of calculating fix-points in a
categorical framework. We first show how fix-points can be expressed
as limit constructions within a suitable category ($\mathcal{C}$ say).
We then transport the fix-point calculation problem to the dual
category of $\mathcal{C}$ ($\mathcal{D}$, say). Within the
dual-category $\mathcal{D}$, the problem gets converted into
calculating a co-limit construction. Still working within the
dual-category, we show that the co-fix-point object can be calculated
using a reachability based iterative algorithm.  Finally the resulting
\emph{dual fix-point} object is transported back across the duality to
$\mathcal{C}$ to give us the fix-point object that we initially wanted
to calculate.

In this paper, we concentrate on the case where the category
$\mathcal{C}$ above, is the category of finite distributive lattices
with homomorphisms, and the dual category $\mathcal{D}$ is the
category of finite partial-orders and monotone functions.

The rest of the paper is structured as follows:
Section~\ref{sec:background} gives some background on lattice theory
and duality-theory for distributive lattices.
Section~\ref{sec:fix-points} formulates the notion of fix-points
within a categorical setting.  Section~\ref{sec:fixp-calc-using} then
transports this calculation to the dual-category, and derives an
algorithm for computing the dual-fix-point object. We conclude with
some comments about future work in Section~\ref{sec:conclusion}.

\section{Background}\label{sec:background}
In this section, we will present some background on lattice theory,
leading up to the duality theory of distributive lattices and
homomorphisms.

\subsection{Lattice Theory}
We assume the basic definitions of lattices and develop only the part
of the theory required for talking about duality.

\begin{definition}[Order Ideal]
  Given a poset $\langle P, \sqsubseteq \rangle$, a subset $S$ of $P$
  is an order ideal if it is closed under the ordering relation of the
  poset.
\[
     x \in S \land y \sqsubseteq x \Rightarrow y \in S
\]
\end{definition}
We will denote the set of order-ideals of a poset, $P$, by
$\mathcal{O}(P)$.

\begin{definition}[Join-irreducible elements]
  Let $L$ be a lattice. An element $x \in L$ is join-irreducible if
  \begin{enumerate}
  \item $x \not = \bot$ and
  \item $x = a \sqcup b$ implies $x = a$ or $x = b$ for all $a,b \in
    L$.
  \end{enumerate}
\end{definition}

Given a lattice $L$, we denote the set of join-irreducible
elements of $L$ by $\mathcal{J}(L)$. The set $\mathcal{J}(L)$ is
also a poset and inherits the order relation of $L$.

\subsection{Duality Theory of Distributive Lattices}
\label{sec:dual-theory-distr}
In this section we present the duality theory of distributive
lattices. Distributive lattices exhibit a duality that is a
generalization of Stone Duality of Boolean algebras. The results
presented in this section are standard results and we refer the reader
to \cite{Davey:2002} for a very fine introduction to this subject.

We begin the section with the definition of distributive lattices and
homomorphisms of distributive lattices.
\begin{definition}[Distributive Lattice]
  Distributive lattices are lattices $\langle L, \sqsubseteq, \sqcap,
  \sqcup, \bot, \top \rangle$ that satisfy the distributive law
\[
    a \sqcap (b \sqcup c) = (a \sqcap b) \sqcup (a \sqcap c) 
\]
for any elements $a$, $b$, and $c$ of the lattice.
\end{definition}

\begin{definition}[homomorphism]
  A function between distributive lattices $f: L \rightarrow K$, is a
  homomorphism if
  \begin{eqnarray*}
    \label{eq:1}
    f(a \sqcap b) = f(a) \sqcap f(b)\\
    f(a \sqcup b) = f(a) \sqcup f(b)\\
    f(\bot) = \bot\\
    f(\top) = \top
  \end{eqnarray*}
\end{definition}

It can be shown that the set of order-ideals of a finite poset
$P$ ordered by subset inclusion, i.e. $\langle \mathcal{O}(P),
\subseteq, \cap, \cup, \emptyset, P \rangle$, forms a
distributive lattice~\cite{Davey:2002}.

\begin{theorem}[Priestley's Theorem (c.f.\ \cite{Davey:2002})]\label{thm:Priestley}
  Let $L$ be a finite distributive lattice. Then the map $\eta : L
  \rightarrow \mathcal{O}(\mathcal{J}(L))$ defined by
\[
   \eta(a) = \{~x \in \mathcal{J}(L)~|~x \sqsubseteq a~\}
\]
is an isomorphism of $L$ onto $\mathcal{O}(\mathcal{J}(L))$
\end{theorem}

\begin{proposition}[Priestley duality(c.f.\
\cite{Davey:2002})]\label{prop:PriestleyDuality}
  Let $P$ and $Q$ be finite posets and let $L = \mathcal{O}(P)$
  and $K=\mathcal{O}(Q)$.
  
  Given a homomorphism $f: L \rightarrow K$, there is an
  associated order-preserving map $\phi_f:Q \rightarrow P$ defined by
  \[
  \phi_f(y) = \text{min}\{~x \in P ~|~ y \in f(\downarrow x)~\}
  \]
  for all $y \in Q$, where $\downarrow x$ is the principal-ideal of $x$.
  
  Given an order-preserving map $\phi:Q \rightarrow P$, there is an
  associated homomorphism $f_\phi : L \rightarrow K$ defined
  as the direct-image of the relation $\phi^{-1}$
  \[
  f_\phi(a) = \phi^{-1}(a)~\text{for all $a \in L$}
  \]
  Equivalently,
  \[
  \phi(y) \in a \Leftrightarrow y \in f_\phi(a) ~ \text{for all $a \in L$, $y \in Q$}
  \]
\end{proposition}

\begin{theorem}[Duality (c.f.\ \cite{Davey:2002})]\label{thm:PriestleyDuality} 
The mappings defined in \Propref{PriestleyDuality} define a
duality between the category ${\mathcal{D}ist}_{fin}$, of finite
distributive lattices and homomorphisms, and the category,
${\mathcal{O}rd}_{fin}$, of finite partial orders and monotone
maps.
\end{theorem}
 \begin{proof}
   The proof of duality between the categories ${\mathcal{O}rd}_{fin}$
   and ${\mathcal{D}ist}_{fin}$ requires establishing two functors
   $\mathcal{O} : {\mathcal{O}rd}_{fin} \rightarrow
   {\mathcal{D}ist}_{fin}^{op}$, and $\mathcal{J} :
   {\mathcal{D}ist}_{fin}^{op} \rightarrow {\mathcal{O}rd}_{fin}$, such
   that $\mathcal{O} \circ \mathcal{J} \cong
   Id_{{\mathcal{D}ist}_{fin}^{op}}$ and $\mathcal{J} \circ \mathcal{O}
   \cong Id_{{\mathcal{O}rd}_{fin}}$. \Thmref{Priestley} and
   \Thmref{PriestleyDuality} provide the functors and isomorphisms that
   allow us to establish the required conditions.
 \end{proof}

\section{Fix-points}\label{sec:fix-points}
In this section, we give an universal characterization of fix-points as
the limit of a particular diagram. This characterization of fix-points
has been inspired by \cite{Jay:TCS:93} which characterizes fix-points
of an endomorphism as an equalizer.

\begin{lemma}
  Given a finite distributive lattice $L$, and an endomorphism $f:L
  \rightarrow L$, the fix-points of $f$ form a distributive lattice.
\end{lemma}
\begin{proof}
   The minimum and maximum fix-points of $f$ are $\bot_L$ and $\top_L$
   respectively for homomorphisms. Given any two fix-points $x$
   and $y$, it is trivial to see that $x \sqcap_L y$ and $x \sqcup_L y$ are
   also fix-points
   \[f(x \sqcup_L y) = f(x) \sqcup_L f(y) = x \sqcup_L y\]
   \[f(x \sqcap_L y) = f(x) \sqcap_L f(y) = x \sqcap_L y\]
   Therefore we can conclude that the lattice of fix-points of $f$ are
   also distributive as any sub-lattice of a distributive lattice is
   distributive \cite{Davey:2002}.
\end{proof}

In categorical language, the category of distributive lattices and
structure preserving maps has \emph{equalizers} (\cite{Jay:TCS:93}) of
the form: \vspace{1em}
\[
\psmatrix[rowsep=1cm]
  F & D & D \\
  F'
\psset{arrows=->,nodesep=2pt}
\ncline[offset=4pt]{1,2}{1,3}^{f}
\ncline[offset=-4pt]{1,2}{1,3}_{id}
\ncline{1,1}{1,2}^{\iota}
\ncline{2,1}{1,2}
\ncline[linestyle=dotted]{2,1}{1,1}
\endpsmatrix
\]
where the distributive lattice $F$ is the lattice of fix-points of the
map $f: D \rightarrow D$, and the homomorphism $\iota$ is an embedding
of the lattice of fix-points, $F$, into $D$.

\section{Fix-point calculation using duality}\label{sec:fixp-calc-using}
The duality between distributive lattices and partial orders given in
the previous section is a very strong result. All \emph{universal}
properties of distributive lattices and homomorphisms can be
equivalently stated and studied as (dual) universal properties of
partial-orders and monotone functions. In particular,
\ref{sec:fix-points} characterizes the fix-points of a homomorphism as
a universal property. By duality we can characterize the fix-points in
in the category ${\mathcal{O}rd}_{fin}$ of partial-orders and monotone
functions as the \emph{co-equalizers}\footnote{The same as the
  equalizer diagram, but with arrow directions reversed.}
\vspace{1em}
\[
\psmatrix[rowsep=1cm]
  C & \mathcal{J}(D) & \mathcal{J}(D) \\
  C'
\psset{arrows=<-,nodesep=2pt}
\ncline[offset=4pt]{1,2}{1,3}^{\mathcal{J}(f)}
\ncline[offset=-4pt]{1,2}{1,3}_{id}
\ncline{1,1}{1,2}^{k}
\ncline{2,1}{1,2}
\ncline[linestyle=dotted]{2,1}{1,1}
\endpsmatrix
\]
Further, this co-equalizer object $C$, is related to the
fix-point object $F$ in ${\mathcal{D}ist}_{fin}$ of $f$, upto
isomorphism by
\begin{equation}
  \label{eq:CoEqualiserFixpointIsomorphism}
   \mathcal{O}(C) \cong F
\end{equation}

Now given $f$ and $g$ of type $A \rightarrow B$ in the category
$\mathcal{S}et$ of sets and functions, their co-equalizer is just a
quotient; i.e.\ the smallest equivalence relation, $\sim$, on $B$,
such that $f(x) \sim g(x)$, for every $x \in A$. We can perform a
similar construction for calculating co-equalizers in
$\mathcal{O}rd_{fin}$ \cite{Goubault:2002}, and build the smallest
preorder $\preceq$ that \emph{extends} the order on $B$, such that
$f(x) \approx g(x)$ for all $x \in A$, where $\approx$ is the
equivalence relation $\preceq \cap \succeq$.  The co-equalizer is then
given by $C \cong B/\approx$. It is easy to see that $C$ is a partial
order, with the order relation, $\sqsubseteq_C$, induced by $\preceq$:

\begin{equation}
  \label{eq:EquivalenceClassOrdering}
    [x]_\approx \sqsubseteq_C [y]_\approx \Leftrightarrow x \preceq y
\end{equation}

\begin{proposition}[c.f.\ \cite{Goubault:2002}]\label{prop:co-equaliser-construction}
  The category, $\mathcal{O}rd_{fin}$, has co-equalizers of $f,g: P
  \rightarrow Q$, given by a partial order whose elements are
  equivalence classes of elements of $Q$.
\end{proposition}

We also observe that since (by construction) $\preceq$ extends
$\sqsubseteq_B$
\begin{equation}
  \label{eq:FixpointOrdering}
  x \sqsubseteq_B y \Rightarrow [x]_\approx \sqsubseteq_C [y]_\approx
\end{equation}

Applying the above construction for our special case of co-equalizer
of $id$ and $\phi_f = \mathcal{J}(f)$, we observe that the
construction boils down to computing the \emph{connected subsets} of
the \emph{map} of $\phi_f$ -- i.e.\ the co-equalizer is the set of
equivalence classes of elements $P$, obtained by identifying $x$,
$\phi_f(x)$, and $\phi_f^{-1}(x)$ for all $x \in P$.

We can now state the relation between the co-equalizer in the category
${\mathcal{O}rd}_{fin}$ and the equalizer in the dual category
$\mathcal{O}(\mathcal{O}rd_{fin})$.
\begin{proposition} \label{prop:CalculatingFixpoints}
  Let $P$ be a finite ordered set and let $L = \mathcal{O}(P)$.
  Given a homomorphism $f: L \rightarrow L$
  \begin{itemize}
  \item {let $\textsf{Fix}(f)$ represent the sub-lattice of $L$,
      consisting of the fix-points of $f$}
  \item {Let $\phi_f : P \rightarrow P$ be as defined in
      \Propref{PriestleyDuality}}
  \item {Let $C = \textsf{Eq}(\phi_f)$, be the co-equalizer of
      $\phi_f$, and $id_P$, calculated as equivalence classes of $P$}
  \item {Let $E = \cup \circ \mathcal{O}(C)$, i.e.\ the distributive
      lattice obtained by calculating the set-union for each
      order-ideal (consisting of sets of equivalence classes) of $C$}
  \end{itemize}

  Now, given any element $X$ of $L$
  \[
  f(X) = X \Leftrightarrow X \in E
  \]
\end{proposition}
\begin{proof}
  \begin{description}
  \item[($\Leftarrow$)] {An order-ideal of $C$, is a set of subsets
      (equivalence classes) of the partial order $P$ whose elements
      are the join-irreducible elements of $L$.  The equivalence
      classes are not themselves down-closed by the ordering on $P$. 
      But it is fairly easy to
      see, from the construction of $C$, that every order-ideal of
      $C$, under $\cup$, gives rise to an order-ideal of $P$.
      
      To show that $f(X) = X$, we observe that $X$ is obtained as a
      union of equivalence classes (of $P$) in $C$. Each of these
      equivalence classes $X_i$ are themselves closed under $\phi_f$
      and $\phi_f^{-1}$ (by the co-limit construction). Therefore
      since $f = f_{\phi_f} = \phi_f^{-1}$
      (\Propref{PriestleyDuality}), each equivalence class
      satisfies the property $f(X_i) = X_i$ and the result follows.}
  \item[($\Rightarrow$)] {Given an order ideal, $X$, such that
      $f(X)=X$. By \Propref{PriestleyDuality}, this implies that
      $\phi_f^{-1}(X) = x$; in other words
      \[
      \forall x \in X \bullet \exists y \in X \bullet y = \phi_f(x)
      \]
      Therefore, since $\phi_f$ is a function, $X$ is closed under $\phi_f$.

      Therefore, in general, $X$ is an union of equivalence classes of
      $P$, that are closed under $\phi_f$ and $\phi_f^{-1}$; i.e.\ $X$
      is an union of elements of $C$ (by the co-equalizer construction
      of \Propref{co-equaliser-construction}). Now,
      \Eqref{FixpointOrdering}, tells us that this set of elements
      of $C$ is an order-ideal in $C$. Therefore $X$ is an element of
      $E$.}
  \end{description}
\end{proof}

\section{Algorithm for fix-point calculation}
\label{sec:alg-dual-fix-point}
Putting all the pieces presented so far, we obtain an algorithm for
calculating fix-points using dual representation of lattices and
homomorphisms.

\begin{algorithm}
\caption{}
\label{alg:CalculatingFixpoints}
\begin{algorithmic}[1]
\REQUIRE $f: L \rightarrow L$
\STATE calculate $P = \mathcal{J}(L)$ and $\phi_f : P \rightarrow P$ (\Propref{PriestleyDuality})
\STATE construct the (undirected) graph $G$ of $\phi_f$ over $P$
\STATE calculate the connected components of $G$, and its ordering based on $P$
\STATE select any order-ideal $M$ of $G$
\STATE \textbf{return} $\bigsqcup_L(\bigcup M)$
\end{algorithmic}
\end{algorithm}

Basically, the algorithm calculates the join-irreducible elements,
$\mathcal{J}(L)$ of a distributive lattice, $L$; and the dual,
$\phi_f$, of a lattice-homomorphism, $f$. It then computes the
connected components, $C$ say, of the un-directed graph of $\phi_f$ on
the elements of $\mathcal{J}(L)$. The set $C$ is partially-ordered and
inherits its ordering from the partial order of $\mathcal{J}(L)$.
Finally, the fix-points of $f$ are computed from the ideals of $C$, as
the least-upper-bound of the sets in a given ideal.

\section{Conclusion and Future Work}
\label{sec:conclusion}
The work presented in this paper is only a starting point for
investigations into alternate algorithms for fix-point calculation.
Future work includes generalizing the framework presented here to
hemimorphisms -- this particular generalization will find immediate
application in the area of data-flow analysis of programs. Another
topic to be further investigated is the connection between the
framework presented here and model-checking algorithms; and relating
the search algorithm of model-checkers over a Kripke-structure to the
equivalence-class calculation performed by the algorithm presented in
this paper.

\bibliographystyle{abbrv}
\bibliography{AbstractInterpretation,CategoryTheory,LatticeTheory,DataFlowFrameworks}
\end{document}